\documentclass[reprint,aps,prd,twocolumn,notitlepage,nofootinbib,showpacs,preprintnumbers,superscriptaddress]{revtex4-1}
\usepackage[T1]{fontenc}
\usepackage[utf8]{inputenc}
\usepackage{bm}
\usepackage{amsmath}
\usepackage{amssymb}
\usepackage{esint}
\usepackage{color}
\usepackage{graphicx}
\PassOptionsToPackage{normalem}{ulem}
\usepackage{ulem}

\usepackage{times}
\usepackage{hyperref}
\usepackage{amsthm,bm,mathrsfs,stmaryrd,amsmath}

\newcommand{\be}{\begin{equation}}
\newcommand{\ee}{\end{equation}}
\newcommand{\ba}{\begin{eqnarray}}
\newcommand{\ea}{\end{eqnarray}}

\newcommand{\nn}{\nonumber\\}

\def\pa{\partial}

\def\a{\alpha}
\def\b{\beta}
\def\g{\gamma}

\def\d{\delta}

\def\l{\lambda}

\def\m{\mu}
\def\n{\nu}

\def\P{\Pi}

\def\r{\rho}

\def\s{\sigma}

\def\o{\omega}

\begin{document}
\preprint{CCQCN-2015-113}

\title{On the absence of the Boulware-Deser ghost in novel graviton kinetic terms}


\author{Wenliang LI}
\email{lii.wenliang@gmail.com}
\affiliation{Crete Center for Theoretical Physics (CCTP) and Crete Center for Quantum Complexity and
Nanotechnology (CCQCN), Department of Physics, University of Crete, P.O. Box 2208, 71003, Heraklion, Greece}


\date{\today}

\begin{abstract}
Using novel nonlinear kinetic terms for gravitons, 
a large class of bi-gravity models were constructed, 
which are potentially free of the Boulware-Deser (BD) ghost.  
In this work, we derive their Hamiltonians using the ADM formalism, 
and verify that the BD ghost is eliminated by additional constraints. 
The general Hamiltonian structure is analogous to 
that of the other class of bi-gravity models free of the BD ghost. 

\end{abstract}


\maketitle

\section{Introduction}
In a general framework for Lagrangian theories  
free of Ostrogradsky's scalar ghost \cite{Li:2015vwa, Li:2015fxa}, 
novel nonlinear kinetic terms for gravitons were proposed in the language of vielbeins
\be
\mathcal L_{\text{kin}} = R\big(E^{(1)}\big)\wedge E^{(2)}\wedge \dots\wedge E^{(d-1)},
\ee
where $d$ is the spacetime dimension, 
$R\big(E^{(1)}\big)$ is the curvature two-form associated with $E^{(1)}$, 
and $E^{(k)}$ could be different vielbeins. 
The Einstein-Hilbert kinetic term corresponds to the case 
where $E^{(k)}$ are the same. 
These nonlinear kinetic terms can be supplemented 
by some nonlinear potential terms 
\cite{deRham:2010kj, Hassan:2011zd, Hinterbichler:2012cn} 
\be
\mathcal L_{\text{pot}} 
=E^{(1)}\wedge \dots\wedge E^{(d)}.
\ee
where $E^{(k)}$ could be different vielbeins.

Some of these novel kinetic terms are nonlinear, multi-gravity completion of 
the two-derivative cubic term discovered in \cite{Folkerts:2011ev}. 
It was shown in \cite{Hinterbichler:2013eza} 
that this cubic term is a natural generalization of 
perturbative Lovelock terms \cite{Lovelock:1971yv} 
and dRGT terms \cite{deRham:2010kj}, 
so more possible terms exist in higher dimensions. 
They were conjectured to have nonlinear completions \cite{deRham:2010kj}, 
which were constructed in \cite{Kimura:2013ika}. 

However, in the literature, there are comprehensive no-go theorems 
for the nonlinear completion of the cubic term mentioned above. 
A no-go theorem for single dynamical metric theories in 4d 
was established in \cite{deRham:2013tfa}: 
around Minkowski vacuum, 
the only nonlinear two-derivative term for spin-2 fields 
that can avoid the 6th degree of freedom 
is the Einstein-Hilbert term.
\footnote{The no-go results for similar constructions in the first-order formulation 
were obtained in a recent work \cite{deRham:2015rxa}. 
Different obstructions were presented in \cite{Matas:2015qxa} as well. } 
The 6th degree of freedom of spin-2 field is a dangerous scalar mode. 
It always plagues a generic nonlinear completion of 
the consistent theory of linear massive gravity, 
namely Fierz-Pauli massive gravity \cite{Fierz:1939ix} .  
This unhealthy scalar mode is known as 
the Boulware-Deser ghost \cite{Boulware:1973my} .

However, our proposals are not ruled out by these negative results, 
because we consider a more general setting where all the spin-2 fields are dynamical
\footnote{In the context of bi-gravity, 
they reduce to the nonlinear proposals 
in \cite{Kimura:2013ika} and \cite{deRham:2013tfa}, 
if we impose symmetric condition \cite{Deser:1974cy} and fix one of the spin-2 fields.}. 
In addition, we use the second-order formulation, 
so the torsion-free condition is satisfied automatically. 

Usually, a bi-gravity model reduces to 
a single dynamical metric model in the decoupling limit 
where one of the Planck masses goes to infinity.   
The novel kinetic terms for bi-gravity will be ruled out 
if they do have the same decoupling limit. 
However, a large class of promising bi-gravity theories 
identified in \cite{Li:2015izu} do not have non-trivial single dynamical metric limit. 
Their Lagrangians contain at most one Einstein-Hilbert term,  
so we have only one Planck mass which can not be sent to infinity. 

Unfortunately, 
precisely due to the fact that only one kind of curvature tensor is allowed, 
the linearized kinetic terms have opposite signs after diagonalization, 
which means one of them is a spin-2 ghost. 
It have been known for a long time that higher derivative gravity  
has the same problem. 
In fact, some of them are equivalent to bi-gravity models with novel kinetic terms 
when the couplings to matter are not introduced \cite{Li:2015izu}. 

According to the minisuperspace analysis in \cite{Li:2015izu} , 
there are two classes of bi-gravity models where the 6th degrees of freedom could be absent. 
In the first class, the kinetic terms are two standard Einstein-Hilbert terms  \cite{Hassan:2011zd}
and novel kinetic terms are not allowed. 
In the second class, the novel graviton kinetic terms could be present. 
We focus on the second class of bi-gravity models in this work. 
In 4d, the possible kinetic terms are
\ba
\mathcal L_1{}^\text{kin} =&\, R(E)\wedge E\wedge E,\label{L1}\\ 
\mathcal L_2{}^\text{kin} =&\, R(E)\wedge E\wedge F,\label{L2}\\
\mathcal L_3{}^\text{kin} =&\, R(E)\wedge F\wedge F\label{L3},
\ea
where $E$ and $F$ are two different vielbeins and 
the normalization factors are not precise. 
If we impose the symmetric condition and fix the second metric to Minkowski, 
they reduce to the two-derivative terms proposed in \cite{Kimura:2013ika}, 
which can also be obtained from 
the dimensionally deconstruction of 
5d Gauss-Bonnet term \cite{deRham:2013tfa}. 
In \cite{Kimura:2013ika, deRham:2013tfa}, 
the second spin-2 field was assumed to be fixed, 
which leads to the dynamical Boulware-Deser ghost. 
Let us emphasize that we will not make this assumption. 

The nonlinear kinetic terms can be accompanied by some potential terms
\ba
\mathcal L_1{}^{\text{pot}}&=&E\wedge E \wedge E\wedge E, 
\label{bi-gr-pot-1}
\\
\mathcal L_2{}^{\text{pot}}&=&E\wedge E \wedge E\wedge F, 
\label{bi-gr-pot-2}
\\
\mathcal L_3{}^{\text{pot}}&=&E\wedge E \wedge F\wedge F, 
\label{bi-gr-pot-3}
\\
\mathcal L_4{}^{\text{pot}}&=&E\wedge F \wedge F\wedge F, 
\label{bi-gr-pot-4}
\\
\mathcal L_5{}^{\text{pot}}&=&F\wedge F \wedge F\wedge F, 
\label{bi-gr-pot-5}
\ea
which are different wedge products of vielbeins $E$ and $F$. 
The cosmological constant terms $\mathcal L_1{}^{\text{pot}}$ and $\mathcal L_5{}^{\text{pot}}$ are special cases 
where only one vielbein is used. 

After a field redefinition, the novel kinetic terms become
\ba
\mathcal L_2{}^\text{kin}&=&
(-)\frac 1 4 \,\sqrt{-g}\,{R(g)_{\m\n}}^{[\m\n}{e_\r}^{\r]}\,d^4x,
\label{L2-r}
\\
\mathcal L_3{}^\text{kin}&=&
 \frac 1 4 \sqrt{-g}\,{R(g)_{\m\n}}^{[\m\n}{e_\r}^{\r}{e_\s}^{\s]}\,d^4x.\label{L3-r}
\ea
where 
$R(g)_{\m\n}{}^{\r\s}$ is the Riemann tensor associated with the metric $g_{\m\n}$
\be
g_{\m\n}=E_\m{}^A E_\n{}^B \eta_{AB}.
\ee
We use the unnormalized convention for the antisymmetrization $[\dots]$ and 
the normalization factors in (\ref{L2-r}, \ref{L3-r}) are precise. 
The new tensor $e_\m{}^\n$ is defined as
\be
{e_\m}^\n={F_\m}^A{(E^{-1})_{A}}^{\n}\label{field-red},
\ee
which coincides with the square root of $g^{\m\r}f_{\r\n}$ 
\be
(\sqrt{g^{-1}f})^\n{}_\m,
\ee
when the symmetric condition is satisfied. 
The second symmetric tensor is defined as
\be
f_{\m\n}=F_\m{}^A F_\n{}^B \eta_{AB}.
\ee
The indices of the second spin-2 field $e$ can be lowered and raised 
by the metric $g_{\m\n}$ and its inverse $g^{\m\n}$. 

This paper is organized as follows. 
In section \ref{sec-ADM}, we introduce the ADM formalism. 
In section \ref{sec-Ham-ex}, we investigate two of the simplest but representative examples. 
In section \ref{sec-Ham-gen}, we discuss the general structure of bi-gravity models 
involving novel kinetic terms. 
In section \ref{sec-concl}, we summarize our results.

\section{ADM formalism}\label{sec-ADM}
Before deriving the Hamiltonians, 
let us express the Lagrangians in terms of the ADM variables 
with the help of Gauss-Coddazzi-Ricci equations . 

In the ADM formalism \cite{Arnowitt:1962hi}, the metric is
\ba
ds^2&=&g_{\m\n}\,dx^\m dx^\n
\nn
&=&-N^2 dt^2
+\g_{ij}(dx^i+N^i dt)
(dx^j+N^j dt)
\nn
&=&
-(n_\m\, dx^\m)^2
+\g_{ij}({\g_\m}^i\, dx^\m)({\g_\n}^j\, dx^\n)
\ea
where 
$N$ is the lapse function, 
$N^i$ is the shift vector, 
$\g_{ij}$ is the induced metric,
$n^\m$ is the normal vector  
and ${\g_\m}^\n$ is the projector, 
according to the foliation of spacetime.

It is useful to introduce a local frame
\be
ds^2=G_{\m\n}\,H^\m H^\n
=- (H^0)^2
+\g_{ij}H^i H^j,
\ee
which we call a local ADM frame. 
This frame generalizes the concept of a "local Lorentz frame".  
The spatial metric 
\be
G_{ij}=\g_{ij}
\ee 
is not necessarily flat 
and coincides with the induced metric $\g_{ij}$. 
The components of the local ADM frame fields are given by
the normal vector $n_\m$ and the projector ${\g_\m}^i$ 
\be
H^0=n_\m \,dx^\m,\quad H^i={\g_\m}^i \,dx^\m.
\ee
For more details about the local ADM frame, 
we refer to section 21.5 of \cite{Misner:1974qy} . 

In the local ADM frame, 
the components of the Riemann curvature tensor 
are given by the Gauss-Codazzi-Ricci relations:
\begin{itemize}
\item
The Gauss equations are
\ba
 R_{ijkl}(G)
 &=&{\g_i}^\m {\g_j}^\n {\g_k}^\r {\g_l}^\s R_{\m\n\r\s}(g)
\nn
&=&R_{ijkl}(\g)+K_{ik}K_{jl}-K_{il}K_{jk}.
\ea
\item
The Peterson-Mainardi-Codazzi equations are
\ba
 R_{ijk0}(G)
& =&
{\g_i}^\m {\g_j}^\n {\g_k}^\r n^\s R_{\m\n\r\s}(g)
\nn
&=&D_i K_{jk}-D_j K_{ik}.
\ea
\item
The Ricci equations are
\ba
 R_{i0k0}(G)
& =&
{\g_i}^\m n^\n {\g_k}^\r n^\s R_{\m\n\r\s}(g)
\nn
&=&-\pounds_n K_{ik}+K_{ij} {K^j}_k+N^{-1} D_i \pa_k N.
\ea
\end{itemize} 
The extrinsic curvature $K_{ij}$ is defined as
\be
K_{ij}=\frac 1 {2}\pounds_n\gamma_{ij}=\frac 1 {2N}(\dot \g_{ij}-D_i N_j-D_j N_i),
\ee
with
\be
n^0=\frac 1 N,\quad n^i=-\frac{N^i}N,
\ee
and $D_i$ is the covariant derivative compatible with the induced metric $\g_{ij}$
\be
D_i \g_{jk}=0.
\ee

Let us introduce a notation
\be
N^0=N, 
\ee
so the lapse function and the shift vector form a spacetime vector
\be
N^\m=(N^0,\,-N^i).
\ee

We use $ R(G)$ to denote the components of the Riemann curvature tensor 
in a local ADM frame.
We can raise the last two indices by $G^{\m\n}$
\ba
{R_{0i}}^{0j}(G)
&=&
R_{0i0k}(G)\,G^{00} G^{kj}
\nn
&=&
\pounds_n {K_i}^j+K_{ik}K^{kj}
-N^{-1}D_i\pa^j N,\label{R_1}
\ea
\ba
{R_{ ij}}^{kl}(G)
&=&
R_{ ijmn}(G)\,G^{mk}G^{nl}
\nn
&=&{R_{ ij}}^{kl}(\g)
+{K_{i}}^k{K_{j}}^l
-{K_{i}}^l{K_{j}}^k,\label{R_2}
\ea
\ba
{R_{jk}}^{0i}(G)
&=&
R_{jk0m}(G)\,G^{00}G^{mi}\qquad\qquad
\nn
&=&
D_j {K_k}^{i}-D_k {K_j}^{i},\label{R_3}
\ea
\ba
{R_{0i}}^{jk}(G)&=&
R_{0imn}(G)\,G^{mj}G^{nk}\qquad\qquad
\nn
&=&
-(D^j {K^k}_{i}-D^k {K^j}_{i}).\label{R_4}
\ea
These equations will be used to derive the expressions of the novel kinetic terms in terms of the ADM variables. 

The components of the new tensor $e_\m{}^\n$ in the local ADM frame are
\ba
e_0{}^0&=&(-){e_\m}^\n n^\m \,n_\n ,
\quad\,
e_0{}^i={e_\m}^\n n^\m \g_\n{}^i,
\nn
e_{ij}&=&{e_\m}^\n  {\g^\m}_i\,\g_{\n j},
\quad
e^i{}_0=e_\m{}^\n  \g^{\m i}\,n_\n,
.
\label{e-ADM}
\ea
The left hand sides $e_0{}^0,\,e_0{}^i,\,e^i{}_0,\,e_{ij}$ 
are the fundamental fields in the local ADM frame. 
Let us emphasis that they are not the components of ${e_\m}^\n$ 
and the indices $0,\,i,\,j,\,k,\dots$ are those of a local ADM frame.

\section{Examples}\label{sec-Ham-ex}
In this section, we carry out the Hamiltonian analyses of 
two concrete examples of novel kinetic terms. 
The discussion of the general structure of 
a linear combination of the nonlinear kinetic terms 
is postponed to the next section. 

\subsection{$\mathcal L_\text{EH}$}
In this subsection, 
we briefly review the Hamiltonian structure of the Einstein-Hilbert kinetic term. 
The general features shared by other kinetic terms are emphasized 
in this well-understood example. 

Using \eqref{R_1} and \eqref{R_2}, we have
\ba
\mathcal L_\text{EH}
&=&\sqrt{-g}\,R(g)
\nn
&=&N\sqrt\g
\big[2\pounds_n {K}+K^2+K_{ij}K^{ij}
+R(\g)
\nn
&&\quad
\qquad-2N^{-1}D_i\pa^i N
\big]
\nn
&=&N\sqrt\g
\left[R(\g)+K_{ij}K^{ij}-K^2\right]
\nn&&
+\pa_\m(N\sqrt\g\, 2n^\m K)-\pa_i(\sqrt\g\,2\pa^i N).\label{L1-expl}
\ea

The Ricci scalar contains a second order time derivative term $\pounds_n K$. 
It can be eliminated by supplementing the action by 
the York-Gibbons-Hawking term 
on the space-like boundaries
\be
S_{YGH}=\int d^3x \big[\sqrt\g\,(-2K)\big]\Big|_{t_i}^{t_f},
\ee
which cancel the time component of the first total derivative term 
in the last line of \eqref{L1-expl}.
The second order space derivative terms can be cancelled 
by terms on the time-like boundaries.

Introducing the matrix
\be
(M_0)^{ijkl}=\g^{ij}\g^{kl}-\g^{ik}\g^{jl},
\label{M0}
\ee
the Lagrangian becomes
\be
\mathcal L_\text{EH}=N\sqrt\g
\big[-K_{ij}(M_0)^{ijkl}K_{kl}+R(\g)\big].
\ee
Note that $M_0$ is the Kulkarni-Nomizu product of two inverse induced metrics 
\be
(M_0)^{ijkl}=\frac 1 2 (\g\owedge\g)^{iljk},
\ee
which is related to the Wheeler-DeWitt metric. 
$(M_0)^{ijkl}$ appears in the novel kinetic terms as well. 

The conjugate momenta are
\be
\Pi_{N}=\frac {\pa \mathcal L_\text{EH}}{\pa N}=0
,\quad
\Pi_{N^i}=\frac {\pa \mathcal L_\text{EH}}{\pa N^i}=0,
\ee
\be
\Pi^{ij}=\frac 1 {2N}\frac {\pa \mathcal L_\text{EH}}{\pa K_{ij}}
=\left(-\frac 1 2\right)\sqrt\g\, (M_0)^{ijkl}\pounds_n\g_{kl},
\ee
where the first line gives primary constraints.

Inverting the relation between $\Pi^{ij}$ and $\pounds_n\g_{kl}$, 
we have
\be
\pounds_n \g_{kl}=(-2)\,\g^{-1/2}\Pi^{ij}(M^{-1}_0)_{ijkl},
\ee
where the inverse matrix of $(M_0)^{ijkl}$ is introduced
\be
(M^{-1}_0)_{ijkl}=\frac 1 2 \g_{ij}\g_{kl}-\g_{ik}\g_{jl},
\label{M0-inv}
\ee
with
\be
(M_0)^{ijkl}(M_0^{-1})_{klmn}=\d_m{}^i\d_n{}^j.
\ee

The Hamiltonian is
\ba
\mathcal H_\text{EH}&=&\dot\g_{ij}\Pi^{ij}-\mathcal L_\text{EH}
\nn
&=&(N\pounds_n \g_{ij}+D_iN_j+D_jN_i)\Pi^{ij}-\mathcal L_\text{EH}
\nn
&=&-N\big[\g^{-1/2}\Pi^{ij}\, (M^{-1}_0)_{ijkl}\,\Pi^{kl}+\sqrt\g\, R\big]
\nn&&
- N^i D_j(2{\Pi_i}^{j}).
\ea

In the second equality, 
we reconstruct $\pounds_n\g_{ij}$ from $\dot \g_{ij}$, 
which generates the $D_i N_j \Pi^{ij}$ terms. 
In the third equality, we express $\pounds_n\g_{ij}$ in terms of $\Pi^{ij}$ using the inverse of $M_0$. 
These are two common steps in 
expressing a Hamiltonian in terms of the canonical variables. 

The momenta are tensor densities, 
so their covariant derivatives should be
\be
D_i{\Pi}^{jk}\rightarrow \sqrt\g\, D_i(\g^{-1/2}{\Pi}^{jk}).
\ee

The total Hamiltonian that contains the information of primary constraints is
\be
\mathcal H_\text{EH}^{\text{T}}=\mathcal H_\text{EH}+\l^\m\, \Pi_{N^\m},
\ee
where $\l^\m$ are Lagrange multipliers associated with the primary constraints. 

To preserve the primary constraints $\Pi_{N^\m}\approx 0$ in time, 
we require the time derivative of $\Pi_{N^\m}$ vanish on the constraint surface
\be
\dot \Pi_{N^\m}=\left\{\Pi_{N^\m}, \int d^3x \,\mathcal H^{\text{T}}_\text{EH}\right\}=(-)\frac {\pa \mathcal H^{\text{T}}_\text{EH}}{\pa N^\m}=(-)\tilde C_\m\approx 0,
\ee
where $A\approx 0$ means $A$ vanishes on the constraint surface. 
They are secondary constraints. 
Since $N^\m$ are Lagrange multipliers,
$\tilde C_\m$ do not contain $N^\m$
\be
\tilde C_0=(-)\left[\g^{-1/2}\Pi^{ij}\, (M^{-1}_0)_{ij,kl}\,\Pi^{kl}+\sqrt\g\, R\right]\approx 0,
\ee
\be
\tilde C_i=(-2)\sqrt\g\, D^j(\g^{-1/2}{\Pi}_{ij})\approx 0,
\ee
where $\tilde C_0\approx 0$ is known as the Hamiltonian constraint 
and $\tilde C_i\approx 0$  the momentum constraint or 
the diffeomorphism constraint. 
There are no more independent constraints 
from the time derivatives of secondary constraints. 

Note that $\tilde C_0$ and $\tilde C_i$ are first-class constraints, 
as their poisson brackets vanish on the constraint surface. 
To see this, we compute the Poisson brackets of 
smeared constraints
\be
\mathcal C_1[\vec \a]=\int d^3 x\, \a^i(x)\,\tilde C_i,\quad
\mathcal C_2[\a]=\int d^3 x\, \a(x)\,\tilde C_0, 
\ee
where $\a(x)$ and $\a^i(x)$ are test functions. 
The result is Dirac's hypersurface deformation algebra
\ba
\{\mathcal C_1[\vec \a],\,\mathcal C_1[\vec \b]\}&=&\mathcal C_1[\pounds_{\vec \a} \vec \b],
\label{scb-gr-1}
\\
\{\mathcal C_1[\vec \a],\,\mathcal C_2[\b]\}&=&\mathcal C_2[\pounds_{\vec \a} \b],
\label{scb-gr-2}
\\
\{\mathcal C_2[\a],\,\mathcal C_2[\b]\}&=&\mathcal C_1[\vec f(\a,\b,\g)],\label{GR-f}
\ea
where boundary terms are neglected,  
the Poisson bracket is defined as
\be
\{A,B\}=\int d^3x\left[
\frac {\d A}{\d \g_{ij}(x)}\frac {\d B}{\d \P^{ij}(x)}
-
\frac {\d B}{\d \g_{ij}(x)}\frac {\d A}{\d \P^{ij}(x)}
\right]
\ee
and the vector in the last bracket \eqref{GR-f} is defined as
\be
f^i(\a,\b,\g^{ij})=\g^{ij}(\a\,\pa_j \b-\b\,\pa_j \a).
\ee 
$f^i$ is also known as structure functions 
\footnote{Therefore, Dirac's hypersurface deformation algebra is not a Lie algebra.}
due to the dependence on the phase space variables $\g_{ij}$. 

The first two brackets (\ref{scb-gr-1}, \ref{scb-gr-2}) can be derived 
in a simple way due to the fact that 
$\mathcal C_1$ generates an orbit in the phase space
\ba
\{\mathcal C_1[\vec \a],\,\g_{ij}\}&=&-\pounds_{\vec \a}\,\g_{ij},\label{GR-diff-1}\\
\{\mathcal C_1[\vec \a],\,\Pi^{ij}\}&=&-\pounds_{\vec \a}\,\Pi^{ij}\label{GR-diff-2},
\ea
which are the Lie derivatives of the canonical variables along $\vec \a$. 
So $\tilde C_i\approx 0$ is also known as 
the diffeomorphism constraint. 
Using \eqref{GR-diff-1} and \eqref{GR-diff-2}, 
the Lie derivative $\pounds_{\vec \a}$ acts on the test functions ($\vec \b$, $\b$) 
after integrating by parts, 
so the first two brackets (\ref{scb-gr-1}, \ref{scb-gr-2}) share a similar form.

\subsection{$\mathcal L_2{}^\text{kin}$}
Let us investigate the simplest example of novel kinetic terms
\be
\mathcal L_2{}^\text{kin}=
(-)\frac 1 4 \,\sqrt{-g}\,{R(g)_{\m\n}}^{[\m\n}{e_\r}^{\r]}\,d^4x.
\label{L2-expl}
\ee 
Since the antisymmetric part of $e_{\m\n}$ 
in $\mathcal L_2{}^\text{kin}$ is projected out, 
we will simply assume $e_{\m\n}$ is symmetric. 
In the local ADM frame, we have
\be
e_0{}^i=e^i{}_0,\quad e_{ij}=e_{ji},.
\ee
Using (\ref{R_1}-\ref{R_4}),
we can derive the explicit expression of $\mathcal L_2{}^\text{kin}$ 
in the local ADM frame 
\ba
\g^{-1/2}\mathcal L_2
&=&
 N (\pounds_n \g_{ij})(\pounds_n \g_{kl})
\Big[{(M_1)}^{ijkl}
-\frac 1 8 {e_0}^{0}(M_0)^{ijkl} \Big]
\nn
&&+ N
(\pounds_n \g_{ij}) (\pounds_n {e_{kl}})\frac 1 2 (M_0)^{ijkl}
\nn&&
+(\pounds_n \g_{ij})\, D_k\big[N{e_0}^{l}(-)(M_0)^{ijk}{}_l\big]
\nn&&
+ NR^{ij}
\Big[(-)e^{kl}(M_0^{-1})_{ijkl} 
+ {e_0}^{0} (-)\frac 1 2 \g_{ij}
\Big]
\nn&&
+ND_i D_j[(M_0)^{ijkl} e_{kl}],
\ea
where $M_0$ and $M_0^{-1}$ are defined in (\ref{M0}, \ref{M0-inv}) and
\be
(M_1)^{ijkl}
=
-\frac 1 2 (\g^{ij}e^{kl}-\g^{ik}e^{jl})
+\frac 1 8 e_m{}^m(M_0)^{ijkl}.
\ee

From the definition in \eqref{e-ADM}, 
we know $e_{ij}$ is tangent to a constant time slice, 
so its Lie derivative is 
\be
\pounds_n e_{ij}=\frac 1 {N}({\dot e}_{ij}-N^k D_k e_{ij}-e_{kj}D_i N^k-e_{ik}D_j N^k).
\ee

We supplement the action by a boundary term 
analogous to the York-Gibbons-Hawking term
\be
\int d^3 x
\big[\sqrt\g\, K_{ij}
(M_0)^{ijkl} e_{kl}
\big]\Big|^{t_f}_{t_i}
\ee
on the space-like boundaries 
to eliminate the second order time derivative term 
in the Lagrangian. 
These boundary terms generate time derivative terms $\pa_t e_{ij}$ in \eqref{L2-expl}, 
so $e_{ij}$ is also a dynamical tensor field. 
In contrast, there is no time derivative acting on ${e_0}^{0},\,e_0{}^i$, 
which are the counterparts of 
the lapse function $N$ and the shift vector $N^i$ 
in the ADM metric. 

The conjugate momenta are
\ba
\Pi^{ij}&=&
\frac {\pa \mathcal L_2}{\pa {\dot\g}_{ij}}
\nn
&=&
2\sqrt{\g}\,(\pounds_n \g_{kl})
\Big[{(M_1)}^{ijkl}
-\frac 1 8 {e_0}^{0}(M_0)^{ijkl} \Big]
\nn
&&+\sqrt{\g}\,\frac 1 2 (\pounds_n {e_{kl}}) (M_0)^{ijkl}
\nn&&
+\sqrt{\g}\,N^{-1}D_k\big[N{e_0}^{l}(-)(M_0)^{ijk}{}_l\big],
\label{PI-2}
\ea
\be
\pi^{ij}=
\frac {\pa \mathcal L_2}{\pa {\dot e}_{ij}}
=\sqrt{\g}\,\frac 1 2 (\pounds_n {\g_{kl}}) (M_0)^{ijkl},
\label{pi-2}
\ee
\be
\Pi_{N^\m}=\frac {\pa \mathcal L_2}{\pa N^\m}=0,
\quad
\pi_{e_0{}^\m}=\frac {\pa \mathcal L_2}{\pa e_0{}^\m}=0,
\ee
where the last line contains 8 primary constraints.

Then we obtain the Hamiltonian by the Legendre transform
\ba
\mathcal H_2
&=&
\dot \g_{ij}\Pi^{ij}+ \dot e_{ij}\pi^{ij}-\mathcal L_2
\nn
&=& \frac 1 2{e_0}^{0}N
\left[\g^{-1/2}\pi ^{ij}\pi^{kl}
 (M_0^{-1})_{ijkl}
+\g^{1/2}R
\right]
\nn&&
+\pi^{ij} D_i\big(2Ne_0{}^l\g_{jl}\big)
\nn&&
+N\g^{-1/2}\pi ^{ij}\pi^{kl}
\Big[-\frac 1 2 e_m{}^m(M_0^{-1})_{ijkl}
\nn&&
\qquad\qquad\qquad\qquad
+\left(e_{ij}\g_{kl}-2\g_{ik}e_{jl}\right)\Big]
\nn&&
+2N\g^{-1/2}\Pi^{ij}\pi^{kl}(M_0^{-1})_{ijkl}
\nn
&&
+N\g^{1/2}{R^{ij}}e^{kl}(M_0^{-1})_{ijkl}
-N\g^{1/2}D_i D_j (M_0 e)^{ij}
\nn
&&
+\Pi^{ij}(D_i N_j+D_j N_i) 
\nn&&
+\pi^{ij}(N^k D_k e_{ij}+e_{kj}D_i N^k+e_{ik}D_j N^k).
\ea

The derivation is similar to the case of $\mathcal H_\text{EH}$. 
We first reconstruct $\pounds_n\g_{ij}, \, \pounds_n e_{ij}$ 
from $\dot\g_{ij},\,\dot e_{ij}$. 
Then, by inverting (\ref{PI-2}, \ref{pi-2}),
we express $\pounds_n\g_{ij}, \, \pounds_n e_{ij}$ 
in terms of $\Pi^{ij},\, \pi^{ij}$. 

The corresponding total Hamiltonian is
\be
\mathcal H_2^{\text{T}}=\mathcal H_2+\l_1^\m\, \Pi_{N^\m}+\l_2^\m\, \Pi_{e_0{}^\m},
\ee
where $\l_1^\m,\,\l_2^\m$ are Lagrange multipliers associated with the primary constraints. 

Note that $N,\,N^i,\,e_0{}^0,\,e_0{}^i$ are Lagrange multipliers in $\mathcal H_2$. 
By computing time derivatives of the primary constraints, 
we obtain 8 secondary constraints which do not contain $N,\,N^i,\,e_0{}^0,\,e_0{}^i$. 
All the secondary constraints are first-class constraints. 
To show this, we introduce the smeared constraints
\ba
\mathcal C_1[\vec \a]&=&
\int d^3 x
\left(\Pi^{ij}\pounds_{\vec \a} \g_{ij} 
+\pi^{ij}\pounds_{\vec \a} e_{ij}\right),
\ea

\ba
\mathcal C_2[\a]=
\int d^3 x\,\a\,\Big\{&&
\g^{-1/2}\pi^{ij}\pi^{kl}
\Big[-\frac 1 2 e_m{}^m(M_0^{-1})_{ijkl}
\nn&&
\qquad\qquad\qquad
+\big(e_{ij}\g_{kl}-2\g_{ik}e_{jl}\big)\Big]
\nn&&
+2\g^{-1/2}\Pi^{ij}\pi^{kl} (M_0^{-1})_{ijkl}
\nn
&&
+\g^{1/2}{R^{ij}}e^{kl}(M_0^{-1})_{ijkl}
\nn&&
-\g^{1/2}D_i D_j \big[(M_0)^{ijkl}{e_{kl}}\big]
\Big\},
\ea

\ba
\mathcal C_3[\vec \a]=\int d^3x\, \left(\pi ^{ij} \pounds_{\vec \a}\g_{ij}\right),
\ea

\be
\mathcal C_4[\a]=\int d^3x\, \a\,[ \g^{-1/2}\pi^{ij}
\pi^{kl}
 {(M_0^{-1})}_{ijkl}
+\g^{1/2}R],
\ee
where $\a(x)$ and $\a^i(x)$ are test functions. 

After a straightforward computation, 
the Poisson brackets of the smeared constraints are
\ba
\{\mathcal C_1[\vec \a],\,\mathcal C_1[\vec \b]\}&=&\mathcal C_1[\pounds_{\vec \a} \vec \b],
\\
\{\mathcal C_1[\vec \a],\,\mathcal C_2[\b]\}&=&
\mathcal C_2[\pounds_{\vec \a} \b],
\\
\{\mathcal C_1[\vec \a],\,\mathcal C_3[\vec \b]\}&=&\mathcal C_3[\pounds_{\vec \a} \vec\b],
\\
\{\mathcal C_1[\vec \a],\,\mathcal C_4[\b]\}&=&\mathcal C_4[\pounds_{\vec \a} \b],
\ea

\ba
\{\mathcal C_2[\a],\,\mathcal C_2[\b]\}&=&
\mathcal C_1[\vec f(\a,\b,\g)]
-\mathcal C_3[\vec f(\a,\b,e)],
\\
\{\mathcal C_3[\vec \a],\,\mathcal C_2[\b]\}&=&\mathcal C_4[\pounds_{\vec \a} \b],
\\
\{\mathcal C_4[\a],\,\mathcal C_2[\b]\}&=&\mathcal C_3[\vec f(\a,\b,\g)],
\ea

\be
\{\mathcal C_3[\vec \a],\,\mathcal C_3[\vec\b]\}
=\{\mathcal C_3[\vec\a],\,\mathcal C_4[\b]\}
=\{\mathcal C_4[\a],\,\mathcal C_4[\b]\}
=0,
\ee
where the structure functions $\vec f$ are defined as
\be
f^i(\a,\b,k^{ij})=k^{ij}(\a\,\pa_j \b-\b\,\pa_j \a).
\ee 
with
\be
k^{ij}=\g^{ij}\quad \text{or }\quad
k^{ij}=e^{ij}=\g^{ik}\g^{jl}e_{kl}.
\ee

The constraints in $\mathcal C_1[\vec \a]$ is 
the bi-gravity version of 
the diffeomorphism constraint. 
The first term in this constraint is the momentum constraint in general relativity,
which generates the orbit of $\g_{ij},\, \Pi_{ij}$ in the phase space.  
The second term in $\mathcal C_1[\vec \a]$ generates the displacement of the second set of canonical variables 
\ba
\{\mathcal C_1(\vec \a),\,e_{ij}\}&=&-\pounds_{\vec \a}\,e_{ij},\\
\{\mathcal C_1(\vec \a),\,\pi^{ij}\}&=&-\pounds_{\vec \a}\,\pi^{ij}.
\ea
Following the same argument as the diffeomorphism constraint in general relativity, 
the Poisson brackets related to $\mathcal C_1$ have the same forms.

Since the Hamiltonian is a linear combination of the secondary constraints
\be
\int d^3 x\,\mathcal H_2=
\mathcal C_1[N^i]+
\mathcal C_2[N]+
\mathcal C_3[N e_0{}^i]+
\mathcal C_4[Ne_0{}^0/2],
\ee 
the time derives of the secondary constraints do not lead to new constraints. 
The primary constraints are of first class as well, 
so we have 16 first class constraints in total. 
We can count the number of dynamical variables
\be
(10+10)\times 2-16\times 2=(2+2)\times 2, 
\ee
where 
a symmetric rank-2 tensor has 10 independent components, 
in the phase space each degree of freedom corresponds to two canonical variables 
and first class constraints eliminate two copies of independent variables. 
Therefore, $\mathcal L_2{}^\text{kin}$ contains $(2+2)$ dynamical degrees of freedom, 
corresponding to two interacting massless gravitons
\footnote{The no-go theorem \cite{Boulanger:2000rq} is evaded because one of the spin-2 kinetic terms has a wrong sign. 
Another way to evade this no-go theorem is to introduce additional fields \cite{Gwak:2015vfb}. 
}. 

\subsection{$\mathcal L_3{}^\text{kin}$} 
The metric $g_{\m\n}$ has only two dynamical degrees of freedom 
because the bi-gravity kinetic terms are covariant. 
In $\mathcal L_2{}^\text{kin}$, 
the 6th degrees of freedom of $e_{\m\n}$ are eliminated by the additional gauge symmetries. 
In this subsection, we show that even though the additional gauge invariances are broken 
in 
\be
\mathcal L_3{}^\text{kin}=
 \frac 1 4 \sqrt{-g}\,{R(g)_{\m\n}}^{[\m\n}{e_\r}^{\r}{e_\s}^{\s]}\,d^4x
\label{L3-expl}
\ee
due to higher order interaction terms, 
there exist two constraints which can eliminate the 6th degree of freedom of the second spin-2 field. 
To minimalize the number of degrees of freedom, 
we impose the symmetric condition \cite{Deser:1974cy}
\be
e_{\m\n}=e_{\n\m},
\ee
as part of the definition of the model. 
In terms of the ADM variables, 
the explicit expression of $\mathcal L_3{}^\text{kin}$ is
\ba
&&\g^{-1/2}\mathcal L_3{}^\text{kin}
\nn
&=&
 N (\pounds_n \g_{ij})(\pounds_n \g_{kl})
\Big[{(M_1)}^{ijkl}
+ {e_0}^{0} {(M_2)}^{ijkl}
\nn&&
\qquad\qquad\qquad\qquad\qquad\quad
+{e_0}^{m}{e_0}^{n} {{(M_3)}^{ijkl}}_{mn}\Big]
\nn
&&+ N
(\pounds_n \g_{ij}) (\pounds_n {e_{kl}}){(M_4)^{ijkl}}
\nn&&
+(\pounds_n \g_{ij})\, D_k\big[N{e_0}^{l}{{(M_5)}^{ijk}}_l\big]
\nn&&
+ N{R_{ij}}
\Big[{(M_6)}^{ij}
+ {e_0}^{0} {(M_7)}^{ij}
+{e_0}^{k}{e_0}^{l} {{(M_8)}^{ij}}_{kl}\Big]
\nn&&
+ND_i D_j (M_9)^{ij},
\ea
where $M_i$ are functions of the dynamical fields $(\g_{ij}, \,e_{ij})$ 
\be
(M_1)^{ijkl}=
-\frac 1 4
\g^{i[j}(m_1)^{k]l}
-\frac 1 2{(m_{1a})}^{ijkl},
\ee
\be
(M_2)^{ijkl}=\frac 1 4
\g^{i[j}(m_{3a})^{k]l},
\ee
\be
(M_3)^{ijkl}{}_{mn}=\frac 1 4
\g^{i[j}(m_{3b})^{k]l}{}_{mn},
\ee
\be
(M_4)^{ijkl}=
(-)\frac 1 2{(m_{1b})}^{ijkl}
,\quad
(M_5)^{ijk}{}_{l}=\frac 1 2
{(m_{2a})}^{i[jk]}{}_l,
\ee
\ba
(M_6)^{ij}
&=&0
,\quad
(M_7)^{ij}=(m_{3a})^{ij},
\ea
\ba
(M_8)^{ij}{}_{kl}= (m_{3b})^{ij}{}_{kl},
\quad
(M_9)^{ij}=-(m_1)^{ij},
\ea
and
\be
(m_1)^{ij}=\g^{ij}{e_k}^{[k}{e_l}^{l]}
-
2e^{i[j}{e_k}^{k]},
\ee
\ba
(m_{1a})^{ijkl}
&=&
(-2)\Big[{e_m}^m
(\g^{ij}e^{kl}
-\g^{ik}e^{jl})
+
e^{ik}{e}^{jl}
\nn&&\quad\quad
+
\g^{ik}e^{jm} e_m{}^l
-
e^{ij}e^{kl}
-
\g^{ij}e^{km}e_m{}^l\Big],
\nn
\ea
\ba
(m_{1b})^{ijkl}&=&
2[{e_m}^m(M_0)^{ijkl}
+
(\g^{ i k}e^{jl}
+
e^{ik}\g^{jl})
\nn&&\qquad\quad
-
(\g^{ij}e^{kl}
+
e^{ij}\g^{ kl})],
\ea
\be
{(m_{2a})}^{kijl}
=(-4)
(\g^{ki}e^{jl}-\g^{ki}\g^{jl}{e_m}^{m}
+
e^{ki}\g^{jl}),
\ee
\be
(m_{3a})^{ij}=
\g^{ij}{e_l}^{l}
-
2e^{ij},
\quad
(m_{3b})^{ijkl}
=2(M_0^{-1})^{ijkl}.
\ee

A boundary term on the space-like boundaries are introduced
\ba
\pa_\m[N\g^{1/2}n^\m K_{ij}
(m_1)^{ij}]
\ea
to cancel the second order time derivative terms. 

We can see the explicit expression of $\mathcal L_3{}^\text{kin}$ 
is considerably more complicated than that of $\mathcal L_2{}^\text{kin}$. 
Nevertheless, we can derive the corresponding Hamiltonian
\ba
\mathcal H_3
&=&
\dot \g_{ij}\Pi^{ij}+ \dot e_{ij}\pi^{ij}-\mathcal L_3
\nn
&=&
(-) N{e_0}^{0}\Big[\g^{-1/2}(\pi M_4^{-1})_{ij}
(\pi M_4^{-1})_{kl}
 {(M_2)}^{ijkl}
 \nn&&\qquad\qquad\qquad\qquad\qquad\qquad
+\g^{1/2}{R_{ij}}
 {(M_7)}^{ij}\Big]
\nn&&
-(\pi M_4^{-1})_{ij} D_k\big[N{e_0}^{l}{{(M_5)}^{ijk}}_l\big]
\nn&&
-N{e_0}^{m}{e_0}^{n}\Big[
\g^{-1/2}(\pi M_4^{-1})_{ij}
(\pi M_4^{-1})_{kl}
 {{(M_3)}^{ijkl}}_{mn}
 \nn&&\qquad\qquad\qquad\qquad\qquad\qquad
+
\g^{1/2}{R_{ij}}{{(M_8)}^{ij}}_{mn}\Big]
\nn&&
-N\g^{-1/2}(\pi M_4^{-1})_{ij}
(\pi M_4^{-1})_{kl}{(M_1)}^{ijkl}
\nn&&
+N\g^{-1/2}\Pi^{ij}(\pi M_4^{-1})_{ij}
-N\g^{1/2}D_i D_j (M_9)^{ij}
\nn
&&
+\Pi^{ij}(D_i N_j+D_j N_i) 
\nn&&
+\pi^{ij}(N^k D_k e_{ij}+e_{kj}D_i N^k+e_{ik}D_j N^k),
\ea
where the conjugate momenta are
\be
\Pi^{ij}=
\frac {\pa \mathcal L_3}{\pa {\dot\g}_{ij}},\quad
\pi^{ij}=
\frac {\pa \mathcal L_3}{\pa {\dot e}_{ij}}
=\sqrt{\g}\, (\pounds_n {\g_{kl}}) (M_4)^{ijkl},
\ee
a shorthand notation is used
\ba
(\pi M_4^{-1})_{mn}=\pi^{ij} (M_4^{-1})_{ijmn},
\ea
and $M_4^{-1}$ is the inverse of $M_4$
\ba
{(M_4)^{ijkl}}(M_4^{-1})_{klmn}=\d_m{}^i\d_n{}^j.
\label{M4-inverse}
\ea

The main difference between the Hamiltonians of 
$\mathcal L_2{}^\text{kin}$ and $\mathcal L_3{}^\text{kin}$ 
is that $e_0{}^i$ are not Lagrange multipliers in the latter case. 
This is due to the fact that $\mathcal L_3{}^\text{kin}$ has less 
gauge symmetries. 
We also expect some secondary constraints of $\mathcal L^{\text{kin}}_3$ are of second class. 

The primary constraints come from the variables that do not have time-derivative terms
\be
\Pi_{N^i}=
\Pi_{N}=
\pi_{e_0{}^i}=
\pi_{{e_0}^{0}}=
0.
\ee

The corresponding total Hamiltonian is
\be
\mathcal H_3^{\text{T}}=\mathcal H_3+\l_1^\m\, \Pi_{N^\m}+\l_2^\m\, \Pi_{e_0{}^\m},
\ee
where $\l_1^\m,\,\l_2^\m$ are Lagrange multipliers associated with the primary constraints. 

Secondary constraints are obtained from the requirement that 
primary constraints are preserved in time. 
In the smeared form, the secondary constraints are
\ba
\mathcal C_1[\vec \a]&=&
\int d^3 x
\left(\Pi^{ij}\pounds_{\vec \a}\, \g_{ij} 
+\pi^{ij}\pounds_{\vec \a}\, e_{ij}\right),
\ea

\ba
\mathcal C_2[ \a]&=&
\int d^3 x\Big\{\frac 1 2
(\pi M_4^{-1})_{ij} 
D_k\big[\a\,{e_0}^{l}{{(M_5)}^{ijk}}_l\big]
\nn&&
+\,\a\,\g^{-1/2}(\pi M_4^{-1})_{ij}
(\pi M_4^{-1})_{kl}{(M_1)}^{ijkl}
\nn&&
-\,\a\,\g^{-1/2}\Pi^{ij}(\pi M_4^{-1})_{ij}
+\,\a\,\g^{1/2}D_i D_j (M_9)^{ij}
\Big\},\nn
\ea

\ba
\mathcal C_3[\vec \a]&=&
\int d^3 x\Big\{
(\pi M_4^{-1})_{ij} D_k\big[\a^l{{(M_5)}^{ijk}}_l\big]
\nn&&
+2\,\a^m{e_0}^{n}\Big[
\g^{-1/2}(\pi M_4^{-1})_{ij}
(\pi M_4^{-1})_{kl}
 {{(M_3)}^{ijkl}}_{mn}
 \nn&&\qquad\qquad\qquad\qquad
+
\g^{1/2}{R_{ij}}{{(M_8)}^{ij}}_{mn}\Big]
\Big\},
\ea

\ba
\mathcal C_4[\a]&=&
\int d^3 x\,
\a\,\Big[\g^{-1/2}(\pi M_4^{-1})_{ij}
(\pi M_4^{-1})_{kl}
 {(M_2)}^{ijkl}
 \nn&&\qquad\qquad\qquad\qquad
+\g^{1/2}{R_{ij}}
 {(M_7)}^{ij}\Big],
\ea
where $\a(x)$ and $\a^i(x)$ are test functions. 

Since $\mathcal C_3$ involves the non-dynamical variables $e_0{}^i$, 
one can in principle solve this equation and 
express $e_0{}^i$ in terms of the dynamical variables $\g_{ij},\,e_{ij},\, \pi^{ij}$
\footnote{Therefore, $e_0{}^i$ are not gauge parameters. 
As $\mathcal C_3$ does not generate gauge symmetry, 
we expect $\mathcal C_3$ is related to second class constraint.}. 
We will not proceed in this way to avoid making $\mathcal H_3$ highly nonlinear in $\pi$ and 
simply think of $\mathcal C_3$ as one of the secondary constraints 
arising from the stability of primary constraints.  

The Hamiltonian is a linear combination of the secondary constraints
\be
\int d^3x\,\mathcal H_3=
\mathcal C_1[N^i]-
\mathcal C_2[N]-
\mathcal C_3[N e_0{}^i/2]-
\mathcal C_4[Ne_0{}^0],
\ee

To examine the existence of tertiary constraints, 
we should compute time derivatives of secondary constraints. 
They are the Poisson brackets of the total Hamiltonian and secondary constraints. 
If they do exist, we should compute the time derivatives of the tertiary constraints 
and repeat the same step  until no independent  constraints are found. 

Since the diagonal diffeomorphism invariance is not broken in $\mathcal L_3{}^\text{kin}$, 
there should be 8 first class constraints 
related to 4 gauge transformations. 
They eliminate most of the dynamical variables in $g_{\m\n}$ 
and only $ 2$ degrees of freedom are propagating dynamically. 

For the components of the second symmetric spin-2 field $e_{\m\n}$, 
we know 4 of them are not dynamical, 
so $e_{\m\n}$ contains at most 6 dynamical degrees of freedom. 
Let us remind the reader that at the linearized level $\mathcal L_3{}^\text{kin}$ 
reduces to two linearized Einstein-Hilbert terms. 
But there are more dynamical degrees of freedom in $e_{\m\n}$ at the nonlinear level  
because the second copy of gauge symmetries are broken by high order interactions. 
This is analogous to massive gravity, 
where the gauge symmetries of the linearized kinetic term are broken by mass terms. 
The difference is that 
the additional gauge symmetries in $\mathcal L_3{}^\text{kin}$ are broken by high order interaction terms, 
rather than mass terms. 

In massive gravity, the 6th degree of freedom is known to be ghost-like, 
which is called the Boulware-Deser ghost \cite{Boulware:1973my} . 
One may suspect that in $\mathcal L_3{}^\text{kin}$ 
the 6th degree of freedom of $e_{\m\n}$ is also dangerous. 
In fact, $\mathcal L_3{}^\text{kin}$ should be supplemented by some mass terms, 
otherwise the helicity-1 modes in $e_{\m\n}$ will become strongly coupled 
due to the lack of kinetic terms. 
Therefore, $\mathcal L_3{}^\text{kin}$ is a natural kinetic term for 
a massive spin-2 field, 
together with a massless one. 
The 6th degree of freedom in $e_{\m\n}$ is closely related to the Boulware-Deser ghost. 

The ghost-like 6th degree of freedom in $\mathcal L_3{}^\text{kin}$ should not be propagating. 
What are the constraints that can eliminate this dangerous degree of freedom?  
The first one can be easily identified with the secondary constraint $\mathcal C_4[\a]$, 
which is generated by the time derivative of the primary constraint $\pi_{{e_0}^{0}}=0$. 

To eliminate the 6th degree of freedom, 
we need one more constraint if $\mathcal C_4[\a]$ is related to a second class constraint. 
\footnote{To count the number of degrees of freedom, 
we need to derive all the independent constraints from 
the stability of primary constraints. 
Then we should compute the Poisson brackets among the constraints
and diagonalize them to determine 
the numbers of first class and second class constraints. 

We will not carry out the full procedure, 
but only show that there are at least two additional constraint equations for the $(10-4)\times 2$ dynamical variables of $e_{\m\n}$. 
The number of independent constraints remains the same after the diagonalization. 
The two additional constraints may be related to first-class or second-class constraints, 
but at least 1 degree of freedom 
in the $6$ dynamical degrees of freedom of $e_{\m\n}$ is eliminated. 
}
In dRGT massive gravity \cite{deRham:2010kj}, 
the BD ghost is eliminated by a secondary constraint and the corresponding tertiary constraint \cite{Hassan:2011hr}. 
We expect that the tertiary constraint generated by the time derivative of $\mathcal C_4[\a]$ 
is the additional constraint we are looking for. 
So our goal is to show that the smeared tertiary constraint
\be
\mathcal C_5[\a]=\{\mathcal C_4[\a],\,\int d^3x\, \mathcal H^{\text{T}}_3(x)\}\approx 0,
\ee
is an independent constraint and does not fix $(N, \,N^i,\,e_0{}^0)$. 
In this way, it eliminates at least one more dynamical variable in the phase space. 

It is straightforward to derive $\mathcal C_5[\a]$. 
The result is long because $M_i$ have complicated dependence on $\g_{ij}$ and $e_{ij}$. 
But the real obstacle is that, 
from the explicit expression of $\mathcal C_5[\a]$, 
we can not immediately figure out 
whether the tertiary constraint is an equation for the dynamical canonical variables,  
which should not determine $N$, $N^i$ or $e_0{}^0$
\footnote{$N$, $N^i$ are gauge parameters, so they should be arbitrary. 
If this constraint fixes $e_0{}^0$, then this is an equation for $e_0{}^0$ 
and does not eliminate the second variable of the 6th degree of freedom.}.

This obstacle stems from the fact that the matrix $M_4^{-1}$ in the Hamiltonian $\mathcal H_3$ 
does not have a closed form expression. 
In particular,  
the canonical momentum $\pi^{ij}$ is always contracted with $M_4^{-1}$
because when we invert the relation between velocity $\dot \g_{ij}$ and momentum $\pi^{ij}$, 
$M_4^{-1}$ is generated, making the obstacle more formidable. 

To derive the result of Poisson brackets, 
we also need to compute the variation of $M_4^{-1}$ with respect to $\g_{ij}$ and $e_{ij}$. 
We make use of the identity below
\be
\d (M_4^{-1})^{ijkl}=(-)(M^{-1}_4)^{ijab} \d(M_4)_{abcd}(M^{-1}_4)^{cd kl}.
\ee
which can be derived from the definition \eqref{M4-inverse} of $M_4^{-1}$ and 
a symmetric property of $M_4$ 
\be
(M_4)_{ijkl}=(M_4)_{klij}.
\ee
Therefore, $\mathcal C_5[\a]$ contains numerous terms involving $M_4^{-1}$ 
and with more complicated index contraction than the terms in the secondary constraints. 
It is not clear how to avoid the unwanted terms
\be
(\a\pa N)(\dots), \,
(N\pa \a)(\dots),
\ee
\be
(\a D\pa N)(\dots),\,
(N D\pa \a)(\dots),
\ee
in the result. For example, some typical unwanted terms are
\be
(\a\pa_i N)\left[\g^{-1/2}\,e_0{}^i \,e_j{}^j\,(M_4^{-1})_{klm}{}^{m}(M_4^{-1}\pi)^{kf}(M_4^{-1}\pi)^l{}_f\right],
\ee
\be
(N\pa_i \a)\left[
\g^{1/2}e_0{}^i e_{jk}(M_4^{-1})^{jklm}R_{lm}\right],
\ee
\be
(\a D_i\pa_j N)\left[e^{ik}\,(M_4^{-1}\pi)_k{}^{j}\right],
\ee
\be
(ND_i\pa_j \a)\left[e_m{}^i \,e_{kl}\,(M_4^{-1})^{jmn}{}_n(M_4^{-1}\pi)^{kl}\right].
\ee

The unwanted terms do not involve $N^i$ and $e_0{}^0$ due to the following brackets
\be
\{\mathcal C_4[\a],\,\mathcal C_1[\vec \b]\}=\mathcal C_4[-\pounds_{\vec \b}\a],
\ee
\be
\{\mathcal C_4[\a],\,\mathcal C_4[\b]\}=0. 
\ee
Using these two brackets, 
the tertiary constraint becomes
\be
\mathcal C_5[\a]=\mathcal C_4[-\pounds_{\vec N}\a]
-\{\mathcal C_4[\a],\,\mathcal C_2[N]+
\mathcal C_3[N e_0{}^i/2]\},
\label{teri-sep}
\ee
where the second term generates 
the unwanted derivative terms of $\a$ and $N$ mentioned above. 

To make one step further, 
we notice that in the single metric limit
\footnote{
This is a trivial single metric limit without a fixed fiducial metric. 
}
\be
e_{ij}\rightarrow \g_{ij}, 
\ee
$M_4^{-1}$ has a closed form expression
\be
(M_4)^{ijkl}\rightarrow (-)(M_0)^{ijkl},
\quad
(M^{-1}_4)^{ijkl}\rightarrow (-)(M^{-1}_0)^{ijkl}.
\ee

In this limit, the second term of \eqref{teri-sep} is simplified
\ba
&&\{\mathcal C_4[\a],\,\mathcal C_2[N]+
\mathcal C_3[N e_0{}^i/2]\}
\big|_{e_{ij}\rightarrow \g_{ij}}
\nn&=&\int d^3x\,
\Big\{
(\a N)(\pi^{ij}+2\Pi^{ij})(R_{ij}-\frac 1 4 \g_{ij}\,R)
\nn&&
+\,(\a N)\Big(\frac 1 2 e_0{}^i e_0{}^j \pi R_{ij}
-\frac 1 2 e_{0i} e_0{}^i \pi R
+e_{0i} e_0{}^i \pi^{jk}R_{jk}
\nn&&
\qquad\qquad
-2 e_0{}^i e_0{}^j \pi_i{}^k R_{jk}
+\frac 3 4 e_0{}^i e_0{}^j \pi_{ij}R\Big)
\nn&&
+2(\a N)\g^{1/2}(R_{ij}-\frac 1 2 R\, \g_{ij})D^i e_0{}^j
\nn&&
+\g^{-1/2}\pa_k(\a N)
\Big[
\Big(\frac 1 8 \pi^2-\frac 1 4 \pi_{ij}\pi^{ij}\Big)e_0{}^k
\nn&&\qquad\qquad\qquad\qquad
-\Big(\frac 1 4 \pi \pi_j{}^k
-\frac 1 2  \pi_{ij}\pi^{ik}\Big)e_0{}^j
\Big]
\nn&&
+\g^{-1}(\a N)
\Big[
-\frac 1 {32}\pi^3
+\frac 3 {16} \pi \pi_{ij}\pi^{ij}
-\frac 1 4 \pi_{ij}\pi^{jk}\pi_k{}^i
\nn&&
\qquad
+e_{0i} \,e_0{}^i\big(-\frac 1 {16} \pi^3+\frac 1 4 \pi \pi_{jk}\pi^{jk}-\frac 1 4 \pi_{jk}\pi^{kl}\pi_l{}^j)
\nn&&
\qquad+e_0{}^i e_0{}^j\Big(\frac 1 2  \pi_{ik}\pi_{jl}\pi^{kl}
-\frac 3 {16}\pi_{ij}\pi_{kl}\pi^{kl}
\nn&&
\qquad\qquad\qquad
-\frac 3 8 \pi_i{}^k\pi_{jk}\pi
+\frac 5 {32}  \pi_{ij}\pi^2
\Big)
\nn&&
\qquad
-\Pi\Big(\frac 1 {16}\pi^2
-\frac 1 8 \pi_{ij}\pi^{ij} \Big)
+\Pi^{ij}\Big(\frac 1 4\pi \pi_{ij}
-\frac 1 2 \pi_{ik}\pi_j{}^k\Big)
\Big]
\nn&&
-\frac 1 2 \mathcal C_3[\g^{ij}(\a\pa_j N-N\pa_j \a)]\Big|_{e_{ij}\rightarrow \g_{ij}}
\Big\},
\ea
where the unwanted terms simply organize into
\be
\frac 1 2 \mathcal C_3[\vec f(\a,N,\g)]\Big|_{e_{ij}\rightarrow \g_{ij}}.
\ee
We want to emphasize that the single metric limit is taken after the Poisson bracket is computed. 

In the single metric limit, 
it is clear that 
\begin{itemize}
\item
$\mathcal C_5[\a]$ is an independent constraint 
which involves cubic momentum terms $ \pi\pi\pi,\,\pi \pi \Pi$, 
\item
and $\mathcal C_5[\a]$ does not fix $N$ in terms of the dynamical variables 
because $N$ is always multiplied by the test function $\a$. 
\end{itemize}
We show that in the subspace of the phase space where $e_{ij}$ and $\g_{ij}$ coincide, 
the tertiary constraint $\mathcal C_5[\a]$ is an equation for the dynamical variable.
Together with the secondary $\mathcal C_4[\a]$,  
the tertiary constraint $\mathcal C_5[\a]$ eliminates the 6th degree of freedom. 

The spirit of the single metric limit is similar to the minisuperspace approximation:  
we consider a special subspace of the phase space 
where the nonlinear structure is considerably simplified and 
tests at the nonlinear level are possible. 
The validities of certain statements in these subspaces are only 
necessary conditions, but non-trivial and beyond the linearized level.

Interestingly, all the unwanted terms are absorbed into the secondary constraint $\mathcal C_3$, 
which might be true beyond the single metric limit. 
To verify this, we make use of the explicit definition of $M_4{}^{-1}$
\ba
&&(-)
[{e_m}^m(M_0)^{ijkl}
+
(\g^{ i k}e^{jl}
+
e^{ik}\g^{jl})
\nn&&\qquad
-
(\g^{ij}e^{kl}
+
e^{ij}\g^{ kl})](M_4{}^{-1})_{klmn}=\d_m{}^i\d_n{}^j, 
\ea
which can reduce the number of $M_4^{-1}$ in $\mathcal C_5[\a]$. 
After a long computation, 
the unwanted terms ($\sim 200$ terms) do reduce to 
\be
-\frac 1 2 \mathcal C_3[\vec f(\a,N,\g)]
\ee 
without taking the single metric limit. 
Therefore, the independent constraint in $\mathcal C_5[\a]$ reads
\ba
\bar{ \mathcal C}_5[\a]&=&\left\{\mathcal C_4[\a/N],\,\int d^3x\, \mathcal H^{\text{T}}_3\right\}
+\mathcal C_4[\pounds_{\vec N}(\a/N)]
\nn&&
-\frac 1 2 \mathcal C_3[\vec f(\a/N,N,\g)]\approx 0.
\ea
The 6th degree of freedom of $e_{\m\n}$ is eliminated 
by 
\be
\mathcal C_4[\a]\approx 0,\quad \bar{ \mathcal C}_5[\a]\approx 0.
\ee 

\section{General situation}\label{sec-Ham-gen}
Now we are ready to discuss the general situation, 
where the Lagrangian is a linear combination of 
three kinetic terms and five potential terms
\ba
\mathcal L
&=&a_1\,\mathcal L_{EH}+a_2\,\mathcal L_2{}^\text{kin}+a_3\,\mathcal L_3{}^\text{kin}
\nn&&
+c_1\,\mathcal L_1{}^\text{pot}
+c_2\,\mathcal L_2{}^\text{pot}
+c_3\,\mathcal L_3{}^\text{pot}
\nn&&
+c_4\,\mathcal L_4{}^\text{pot}
+c_5\,\mathcal L_5{}^\text{pot},
\ea
and at least one of the coefficients of the novel kinetic term is not zero
\be
a_2\neq 0
\quad \text{or}\quad a_3\neq 0. 
\ee

As we discuss below, there are 2 types of bi-gravity models, 
which extend the main features of the two examples 
$\mathcal L=\mathcal L_2{}^\text{kin}$ and 
$\mathcal L=\mathcal L_3{}^\text{kin}$. 

The definitions of the kinetic terms are the same as those in 
the examples
\ba
\mathcal L_\text{EH}&=&\sqrt{-g}\,R(g),
\\
\mathcal L_2{}^\text{kin}&=&
(-)\frac 1 4 \,\sqrt{-g}\,{R(g)_{\m\n}}^{[\m\n}{e_\r}^{\r]}\,d^4x,
\\
\mathcal L_3{}^\text{kin}&=&
 \frac 1 4 \sqrt{-g}\,{R(g)_{\m\n}}^{[\m\n}{e_\r}^{\r}{e_\s}^{\s]}\,d^4x,
\ea
and the potential terms (\ref{bi-gr-pot-1}-\ref{bi-gr-pot-5}) are
\ba
\mathcal L_1{}^{\text{pot}}&=&\sqrt{-g}\,, 
\\
\mathcal L_2{}^{\text{pot}}&=&\sqrt{-g}\,e_\m{}^\m, 
\\
\mathcal L_3{}^{\text{pot}}&=&\sqrt{-g}\,e_\m{}^{[\m} e_\n{}^{\n]}, 
\\
\mathcal L_4{}^{\text{pot}}&=&\sqrt{-g}\,e_\m{}^{[\m} e_\n{}^{\n}e_\r{}^{\r]}, 
\\
\mathcal L_5{}^{\text{pot}}&=&\sqrt{-g}\,e_\m{}^{[\m} e_\n{}^{\n}e_\r{}^{\r}e_\s{}^{\s]}.
\ea

In the local ADM frame, 
the explicit expression of the general Lagrangian is
\ba
\g^{-1/2}\mathcal L
&=&
 N (\pounds_n \g_{ij})(\pounds_n \g_{kl})
\Big[{(M_1)}^{ijkl}
+ {e_0}^{0} {(M_2)}^{ijkl}
\nn&&
\qquad\qquad\qquad\qquad\qquad\quad
+{e_0}^{m}{e_0}^{n} {{(M_3)}^{ijkl}}_{mn}\Big]
\nn
&&+ N
(\pounds_n \g_{ij}) (\pounds_n {e_{kl}}){(M_4)^{ijkl}}
\nn&&
+(\pounds_n \g_{ij})\, D_k\big[N{e_0}^{l}{{(M_5)}^{ijk}}_l\big]
\nn&&
+ N{R_{ij}}
\Big[{(M_6)}^{ij}
+ {e_0}^{0} {(M_7)}^{ij}
+{e_0}^{k}{e_0}^{l} {{(M_8)}^{ij}}_{kl}\Big]
\nn\\&&
+ND_i D_j (M_9)^{ij}
\nn&&
+N\left[
e_0{}^0\,M_{10}
+e_0{}^ie_0{}^j\,(M_{11})_{ij}
+M_{12}
\right],
\ea
which is explicitly linear in $N,\,N^i,\, e_0{}^0$. 

The matrices $M_i$ are functions of the dynamical fields $(\g_{ij}, \,e_{ij})$ 
\be
(M_1)^{ijkl}=
-\frac 1 4
\g^{i[j}(m_1)^{k]l}
-\frac 1 2{(m_{1a})}^{ijkl}
+\frac 1 4
\g^{i[j}(m_{3c})^{k]l},
\ee
\be
(M_2)^{ijkl}=\frac 1 4
\g^{i[j}(m_{3a})^{k]l},
\ee
\be
(M_3)^{ijkl}{}_{mn}=\frac 1 4
\g^{i[j}(m_{3b})^{k]l}{}_{mn},
\ee
\be
(M_4)^{ijkl}=
(-)\frac 1 2{(m_{1b})}^{ijkl}
,\quad
(M_5)^{ijk}{}_{l}=\frac 1 2
{(m_{2a})}^{i[jk]}{}_l,
\ee
\ba
(M_6)^{ij}
&=&(m_{3c})^{ij}
,\quad
(M_7)^{ij}=(m_{3a})^{ij},
\ea
\ba
(M_8)^{ij}{}_{kl}= (m_{3b})^{ij}{}_{kl},
\quad
(M_9)^{ij}=-(m_1)^{ij},
\ea
\ba
M_{10}=\sum_{n=2}^5 c_n \, (n-1)\, e_{i_1}{}^{[i_1}\dots e_{i_{n-2}}{}^{i_{n-2}]},
\ea
\be
(M_{11})_{kl}=\sum_{n=3}^5 c_n \, (n-1)(n-2)\,
e_{[i_1}{}^{i_1}\dots e_{i_{n-3}}{}^{i_{n-3}}\g_{k]l},
\ee
\ba
M_{12}=\sum_{n=1}^4 c_n\, e_{i_1}{}^{[i_1}\dots e_{i_{n-1}}{}^{i_{n-1}]},
\ea
and
\ba
(m_1)^{ij}&=&2a_1\g^{ij}-a_2(M_0)^{ijkl}e_{kl}
\nn&&
+a_3(\g^{ij}{e_k}^{[k}{e_l}^{l]}
-
2e^{i[j}{e_k}^{k]}),
\ea
\ba
(m_{1a})^{ijkl}
&=&a_2(\g^{ij}e^{kl}-\g^{ik}e^{jl})
\nn&&
-2a_3\Big[{e_m}^m
(\g^{ij}e^{kl}
-\g^{ik}e^{jl})
+
e^{ik}{e}^{jl}
\nn&&\quad\quad
+
\g^{ik}e^{jm} e_m{}^l
-
e^{ij}e^{kl}
-
\g^{ij}e^{km}e_m{}^l\Big],
\nn
\ea
\ba
(m_{1b})^{ijkl}&=&
(-a_2)(M_0)^{ijkl}
\nn&&
2a_3[{e_m}^m(M_0)^{ijkl}
+
(\g^{ i k}e^{jl}
+
e^{ik}\g^{jl})
\nn&&\qquad\quad
-
(\g^{ij}e^{kl}
+
e^{ij}\g^{ kl})],
\ea
\ba
{(m_{2a})}^{kijl}
&=&
(-a_2)(M_0)^{ijkl}
\nn&&
-4a_3
(\g^{ki}e^{jl}-\g^{ki}\g^{jl}{e_m}^{m}
+
e^{ki}\g^{jl}),
\quad
\ea
\be
(m_{3a})^{ij}=
-\frac 1 2 a_2\g^{ij}
+
2a_3(M_0^{-1})^{ijkl}e_{kl},
\ee
\be
(m_{3b})^{ijkl}
=2a_3(M_0^{-1})^{ijkl},
\ee
\be
(m_{3c})^{ij}
=a_1\g^{ij}-a_2(M_0^{-1})^{ijkl}e_{kl},
\ee

The original Lagrangian is supplemented by boundary terms on the space-like boundaries
\ba
\pa_\m[N\g^{1/2}n^\m K_{ij}
(m_1)^{ij}]
\ea
to cancel the second order time derivative terms. 
They generalize the York-Gibbons-Hawking term. 

The Hamiltonian is derived by the Legendre transform
\ba
\mathcal H
&=&
\dot \g_{ij}\Pi^{ij}+ \dot e_{ij}\pi^{ij}-\mathcal L
\nn
&=&(-) N{e_0}^{0}\Big[\g^{-1/2}(\pi M_4^{-1})_{ij}
(\pi M_4^{-1})_{kl}
 {(M_2)}^{ijkl}
 \nn&&\qquad\qquad\qquad\qquad
+\g^{1/2}{R_{ij}}
 {(M_7)}^{ij}
+M_{10}
\Big]
\nn&&
-(\pi M_4^{-1})_{ij} D_k\big[N{e_0}^{l}{{(M_5)}^{ijk}}_l\big]
\nn&&
-N{e_0}^{m}{e_0}^{n}\Big[
\g^{-1/2}(\pi M_4^{-1})_{ij}
(\pi M_4^{-1})_{kl}
 {{(M_3)}^{ijkl}}_{mn}
 \nn&&\qquad\qquad
+
\g^{1/2}{R_{ij}}{{(M_8)}^{ij}}_{mn}
+\g^{1/2}(M_{11})_{mn}\Big]
\nn&&
-N\g^{-1/2}(\pi M_4^{-1})_{ij}
(\pi M_4^{-1})_{kl}{(M_1)}^{ijkl}
\nn&&
+N\g^{-1/2}\Pi^{ij}(\pi M_4^{-1})_{ij}
- N\g^{1/2}{R_{ij}}
{(M_6)}^{ij}
\nn&&
-N\g^{1/2}D_i D_j (M_9)^{ij}
-N\g^{1/2}M_{12}
\nn
&&
+\Pi^{ij}(D_i N_j+D_j N_i) 
\nn&&
+\pi^{ij}(N^k D_k e_{ij}+e_{kj}D_i N^k+e_{ik}D_j N^k),
\ea
where a shorthand notation is used
\ba
(\pi M_4^{-1})_{mn}=\pi^{ij} (M_4^{-1})_{ijmn},
\ea
and $M_4^{-1}$ is the inverse of $M_4$
\ba
{(M_4)^{ijkl}}(M_4^{-1})_{klmn}=\d_m{}^i\d_n{}^j.
\ea

The conjugate momenta are defined as
\be
\Pi^{ij}=
\frac {\pa \mathcal L}{\pa {\dot\g}_{ij}}
,\quad
\pi^{ij}=
\frac {\pa \mathcal L}{\pa {\dot e}_{ij}}
=\sqrt{\g}\, (\pounds_n {\g_{kl}}) (M_4)^{ijkl},
\ee
\be
\Pi_{N^\m}=\frac {\pa \mathcal L}{\pa N^\m}=0,
\quad
\pi_{e_0{}^\m}=\frac {\pa \mathcal L}{\pa e_0{}^\m}=0,
\ee
where the last line contains 8 primary constraints.

To encode the information of primary constraints, 
we introduce the total Hamiltonian
\be
\mathcal H^{\text{T}}=\mathcal H+\l_1^\m\, \Pi_{N^\m}+\l_2^\m\, \Pi_{e_0{}^\m},
\ee
where $\l_1^\m,\,\l_2^\m$ are Lagrange multipliers associated with 8 primary constraints. 

The total Hamiltonian is a linear combination of constraints, 
so it vanishes on the constraint surface
\be
\mathcal H^T\approx 0. 
\ee

According to their Hamiltonian structures, 
the 4d bi-gravity models involving novel kinetic terms are classified into two types
\footnote{This classification can be easily generalized to higher dimensions, 
where more derivative terms are allowed. 
In Type A models, there is at most one $F$ vielbein in the wedge products. }
:
\begin{itemize}
\item
Type A
\be
a_2\neq 0,\quad a_3=c_3=c_4=c_5=0,
\ee
where the number of dynamical degrees of freedom is
\be
\# (\text{Type A}) = 2 + 2 .
\ee
\item
Type B
\be
\text{at least one of}\quad
a_3,\,c_3,\,c_4,\,c_5\neq 0,
\ee
where the number of dynamical degrees of freedom is
\be
\#(\text{Type B})
\le 2 + 5 .
\ee
\end{itemize}

The representative examples are 
$\mathcal L_2{}^\text{kin}$ and $\mathcal L_3{}^\text{kin}$ 
discussed in the previous section, 
which respectively capture the characteristic features of these two types of models.

\subsection{Type A}
Type A models describe two interacting, gauge invariant, massless spin-2 fields. 
Both the primary and the secondary constraints are of first-class. 
There are no more independent constraints. 
Let us define the smeared secondary constraint as
\ba
\mathcal C_1[\vec\a]&=&\frac 1 {a_2}\int d^3x\, \a^i(x)\frac {\d}{\d N^i(x)}\int d^3 y\, \mathcal H(y),
\\
\mathcal C_2[\a]&=&\frac 1 {a_2}\int d^3x\, \a(x)\frac {\d}{\d N(x)}\int d^3 y\, \mathcal H(y)
\nn&&
-\mathcal C_3[N e_0{}^i]-\mathcal C_4[N e_0{}^0 /2],
\\
\mathcal C_3[\vec\a]&=&\frac 1 {a_2}\int d^3x\, \frac{\a^i(x)}{N(x)}\frac {\d}{\d e_0{}^i(x)}\int d^3 y\, \mathcal H(y),
\\
\mathcal C_4[\a]&=&\frac 1 {a_2}\int d^3x\, \frac{2\a(x)}{N(x)}\frac {\d}{\d e_0{}^0}\int d^3 y\, \mathcal H(y)
\ea

Their Poisson brackets are the same as those in the case of $\mathcal L_2{}^\text{kin}$ 
\ba
\{\mathcal C_1[\vec \a],\,\mathcal C_1[\vec \b]\}&=&\mathcal C_1[\pounds_{\vec \a} \vec \b],
\\
\{\mathcal C_1[\vec \a],\,\mathcal C_2[\b]\}&=&
\mathcal C_2[\pounds_{\vec \a} \b],
\\
\{\mathcal C_1[\vec \a],\,\mathcal C_3[\vec \b]\}&=&\mathcal C_3[\pounds_{\vec \a} \vec\b],
\\
\{\mathcal C_1[\vec \a],\,\mathcal C_4[\b]\}&=&\mathcal C_4[\pounds_{\vec \a} \b],
\ea

\ba
\{\mathcal C_2[\a],\,\mathcal C_2[\b]\}&=&
\mathcal C_1[\vec f(\a,\b,\g)]
-\mathcal C_3[\vec f(\a,\b,e)],
\\
\{\mathcal C_3[\vec \a],\,\mathcal C_2[\b]\}&=&\mathcal C_4[\pounds_{\vec \a} \b],
\\
\{\mathcal C_4[\a],\,\mathcal C_2[\b]\}&=&\mathcal C_3[\vec f(\a,\b,\g)],\label{gen-Poi-42}
\ea

\be
\{\mathcal C_3[\vec \a],\,\mathcal C_3[\vec\b]\}
=\{\mathcal C_3[\vec\a],\,\mathcal C_4[\b]\}
=\{\mathcal C_4[\a],\,\mathcal C_4[\b]\}
=0,
\ee
where the structure functions are given by
\be
f^i(\a,\b,k^{ij})=k^{ij}(\a\,\pa_j \b-\b\,\pa_j \a).
\ee
The Poisson brackets involving primary constraints vanish. 
Therefore, both the primary and the secondary constraints 
are first class constraints.
\\

Type A models have $(2+2)$ dynamical degrees of freedom 
corresponding to two massless gravitons.  
The gauge symmetries are
\begin{itemize}
\item
diagonal diffeomorphism invariance
\be
\d g_{\m\n}=\pounds_\xi g_{\m\n},\quad
\d e_{\m\n}=\pounds_\xi e_{\m\n}, 
\ee
\item
additional ``diffeomorphism invariance"
\be
\d e_{\m\n}=\pounds_{\xi'} g_{\m\n},
\ee

\item
diagonal local Lorentz invariance
\be
\d E_\m{}^A= \o_B{}^A\, E_\m{}^B,
\quad
\d F_\m{}^A= \o_B{}^A\, F_\m{}^B,
\ee

\item
additional ``local Lorentz invariance"
\be
\d F_\m{}^A=\o'_B{}^A\, E_\m{}^B, 
\ee
\end{itemize}
where $\xi^\m,\,\xi'^\m$ are four vectors and $\o_B{}^A,\,\o'_B{}^A$ are antisymmetric. 
At the linear level, the two diffeomorphism invariances reduce to 
two sets of linearized symmetries of the Lagrangians, 
which consists of two decoupled linearized Einstein-Hilbert kinetic terms.

In the minisuperspace approximation, 
the right hand side of 
\be
\{\mathcal C_4[\a],\,\mathcal C_2[\b]\}=\mathcal C_3[\vec f(\a,\b,\g)]
\ee
vanishes. 
It was speculated in \cite{Li:2015izu} that 
the two commuting Hamiltonian-like constraints are of first-class. 
Here we can see this statement is indeed true 
and we have two sets of first class constraints. 

The nonlinear models in Type A have the same amount of 
dynamical degrees of freedom 
as the linearized theories, 
which contain two free massless spin-2 fields 
whose kinetic terms have opposite signs. 

\subsection{Type B}
In Type B models \footnote{For simplicity, we assume the symmetric condition \cite{Deser:1974cy} for $e_{\m\n}$.}, 
the additional gauge symmetries are broken 
by the derivative interaction of $\mathcal L_3{}^\text{kin}$ 
or the potential interactions $\mathcal L_3{}^\text{pot},\,\mathcal L_4{}^\text{pot},\,\mathcal L_5{}^\text{pot}$. 
A direct consequence is that $e_0{}^i$ are not Lagrange multipliers in the Hamiltonian.  

The smeared secondary constraints of Type B models are
\ba
\mathcal C_1[\vec \a]&=&
\int d^3 x
\left(\Pi^{ij}\pounds_{\vec \a}\, \g_{ij} 
+\pi^{ij}\pounds_{\vec \a}\, e_{ij}\right),
\ea

\ba
\mathcal C_2[ \a]&=&
\int d^3 x\Big\{\frac 1 2
(\pi M_4^{-1})_{ij} 
D_k\big[\a\,{e_0}^{l}{{(M_5)}^{ijk}}_l\big]
\nn&&
+\,\a\,\g^{-1/2}(\pi M_4^{-1})_{ij}
(\pi M_4^{-1})_{kl}{(M_1)}^{ijkl}
\nn&&
-\,\a\,\g^{-1/2}\Pi^{ij}(\pi M_4^{-1})_{ij}
+\,\a\,\g^{1/2}R_{ij}(M_6)^{ij}
\nn
&&
+\,\a\,\g^{1/2}D_i D_j (M_9)^{ij}
+\a \g^{1/2}M_{12}
\Big\},
\ea

\ba
\mathcal C_3[\vec \a]&=&
\int d^3 x\Big\{
(\pi M_4^{-1})_{ij} D_k\big[\a^l{{(M_5)}^{ijk}}_l\big]
\nn&&
+2\,\a^m{e_0}^{n}\big[
\g^{-1/2}(\pi M_4^{-1})_{ij}
(\pi M_4^{-1})_{kl}
 {{(M_3)}^{ijkl}}_{mn}
 \nn&&\qquad
+
\g^{1/2}{R_{ij}}{{(M_8)}^{ij}}_{mn}
+\g^{1/2}(M_{11})_{mn}
\big]\Big\},
\ea

\ba
\mathcal C_4[\a]&=&
\int d^3 x\,
\a\,\Big[\g^{-1/2}(\pi M_4^{-1})_{ij}
(\pi M_4^{-1})_{kl}
 {(M_2)}^{ijkl}
 \nn&&\qquad\qquad
+\g^{1/2}{R_{ij}}
 {(M_7)}^{ij}
 +\g^{1/2}M_{10}\Big].
\ea

Then the Hamiltonian is a linear combination of the smeared secondary constraints
\be
\int d^3x\,\mathcal H=
\mathcal C_1[N^i]-
\mathcal C_2[N]-
\mathcal C_3[N e_0{}^i/2]-
\mathcal C_4[Ne_0{}^0].
\ee

Now let us examine the smeared tertiary constraint
\be
\mathcal C_5[\a]=\{\mathcal C_4[\a],\,\int d^3x\, \mathcal H^{\text{T}}(x)\}\approx 0.
\ee
We expect that 
$\mathcal C_5[\a]$ is an independent constraint for the dynamical variables. 

According to the definition of $\mathcal C_5[\a]$, 
we compute the complete Poisson brackets, which is a lengthy result. 
When $a_3\neq 0$, $\mathcal C_5[\a]$ contains cubic momentum terms, 
so it cannot be written as a linear combination of the secondary constraints. 
When $a_3= 0$, no cubic momentum term appears in $\mathcal C_5[\a]$, 
but there is no apparent way to express $\mathcal C_5[\a]$ in terms of other constraints. 
We conclude that $\mathcal C_5[\a]$ is an independent constraint. 

As before, $\mathcal C_5[\a]$ contains a lot of unwanted terms
\be
(\a\pa N)(\dots), \,
(N\pa \a)(\dots),
\ee
\be
(\a D\pa N)(\dots),\,
(N D\pa \a)(\dots), 
\ee
but we expect them to be absorbed into
\footnote{When the coefficients of the symmetry breaking terms vanish, 
Type B models becomes Type A models. 
The presence of $\mathcal C_3$ in $\mathcal C_5$ is consistent with \eqref{gen-Poi-42}. } 
\be
\frac 1 2 \mathcal C_3[\vec f(\a,N,\g)].
\ee
It is difficult to directly check this statement in the general case 
due to a proliferation of terms in $\mathcal C_5[\a]$. 
However, in the single metric limit $e_{ij}\rightarrow \g_{ij}$, 
the unwanted terms are greatly simplified.  
In this limit, we are able to verify that they do organize into $\mathcal C_3$. 
\\

Here we would like to discuss some general properties of the diffeomorphism constraint. 
We can see the secondary constraints in $\mathcal C_3$ determine $e_0{}^i$ in terms of the canonical variables $\g_{ij},\, e_{ij},\,\pi^{ij}$. 
In addition, the quaternary constraint
\be
\mathcal C_6[\a]=\{\mathcal C_5[\a],\,\int d^3x\, \mathcal H^{\text{T}}(x)\}\approx 0
\ee
is usually an equation for $e_0{}^0$ 
because the Poisson bracket of $\mathcal C_4$ and $\mathcal C_5$ does not vanish
\footnote{
In Weyl gravity, 
the Poisson bracket $\{\mathcal C_5[\a],\,\mathcal C_4[\b]\}$ vanishes, 
so $e_0{}^0$ remains arbitrary. 
The diffeomorphism constraint should not contain 
\be
\pi_{e_0{}^0}\pounds_{\vec \a}\,e_0{}^0.
\ee
}. 
Since $e_0{}^0$, $e_0{}^i$ are not arbitrary functions, 
the diffeomorphism constraint
\footnote{The diffeomorphism constraint corresponds to some first class constraints 
which can be identified after the constraint brackets are diagonalized. 
} 
is modified such that its Poisson brackets with other constraints only change the test functions.  
In the smeared form, the diffeomorphism constraint should be
\be
\mathcal C^{\text{diff}}[\vec \a]=\mathcal C_1[\vec \a]
+\int d^3 x\, (\pi_{e_0{}^0}\pounds_{\vec \a}\,e_0{}^0
+\pi_{e_0{}^i}\pounds_{\vec \a}\,e_0{}^i).
\ee
Then $\mathcal C^{\text{diff}}[\a]$ generates an orbit in the phase space
\be
\{\mathcal C_1[\vec \a],\,\mathcal A\}=-\pounds_{\vec \a}\,\mathcal A,
\ee
where $\mathcal A$ can be the canonical variables
\be
\mathcal A=\g_{ij},\, e_{ij},\,e_0{}^0,\,e_0{}^i,\,\Pi^{ij},\, \pi^{ij},\,\pi_{e_0{}^0},\,\pi_{e_0{}^i},
\ee
but not the gauge parameters and their conjugate momenta. 
The changes are the Lie derivatives of the canonical variables along $\vec \a$. 
After integrating by parts, we have
\be
\{\mathcal C^{\text{diff}}[\vec \a],\, C_i[\vec \b]\}=C_i[\pounds_{\vec \a}\vec \b],\quad i=1,3
\ee
and
\be
\{\mathcal C^{\text{diff}}[\vec \a],\, C_i[\b]\}=C_i[\pounds_{\vec \a}\b],\quad i=2,4.
\ee
Therefore, the linear combinations of the primary and secondary constraints in 
$\mathcal C^{\text{diff}}[\vec \a]$ are first class constraints. 
This is analogous to the Hamiltonian structure of spatially covariant gravity \cite{BH-ham}, 
where one of the first class constraints is identified with 
a linear combination of 
the momentum constraint and the lapse primary constraint. 
\\

Now we continue the counting of dynamical degrees of freedom. 
We have two constraint equations $\mathcal C_4$ and $\mathcal C_5$ 
for $(2+6)\times 2$ dynamical variables. 
If they are related to second class constraints, 
there are at most $(2+5)$ degrees of freedom 
because a second class constraint removes $1/2$ degree of freedom. 
If $\mathcal C_5$ is related to a first class constraint, 
the number of dynamical degrees of freedom will be at most $(2+9/2)$ 
as a first class constraint eliminates 1 degree of freedom. 
Usually, a first class constraint is related to a gauge symmetry, 
then the arbitrary gauge parameter will not appear in the constraints. 
The primary constraint of the gauge parameter should be a first class constraint 
and there are at most $(2+4)$ dynamical degrees of freedom. 

A well-known example is Weyl gravity
\be
\mathcal L_\text{Weyl}=R(E)\wedge E\wedge F+E\wedge E\wedge F \wedge F,
\ee  
which is conformal invariant and contains $(2+4)$ degrees of freedom. 
The independent part of the tertiary constraint 
\be
\bar{ \mathcal C}_5[\a]=\int d^3x\,\left(
2\Pi_i{}^i- \pi^{ij}D_i\pa_j
\right)\a
\ee 
is related to the generator of conformal transformations 
\be
\d \g_{\m\n}=2 \a\, \g_{\m\n},\quad \d e_{\m\n}=-\nabla_\m\pa_\n \a.
\ee 
The non-dynamical variable $e_0{}^0$ remains arbitrary. 
After diagonalizing the constraint brackets, 
$\mathcal C_5$ is supplemented by some terms involving $\pi_{e_0{}^i}$ 
in order to preserve the equation for $e_0{}^i$ under a conformal transformation. 
Then we have one more pair of first class constraints. 
Note that Weyl gravity is also a nonlinear completion of 
Fierz-Pauli theory at the partially massless points of the parameter space. 

\section{conclusion}\label{sec-concl}
In summary, we show that the BD ghost can be removed 
by additional constraints in the bi-gravity models with novel kinetic terms. 
There are two important features of the general Hamiltonian structure 
\footnote{These two features originate in the antisymmetric structures 
in the Lagrangians and can be easily generalized to other dimensions.}:
\begin{itemize}
\item
The first key point is that the Hamiltonian is always linear in $e_0{}^0$, 
so the time derivative of the primary constraint 
$\pi_{e_0{}^0}=0$  
generates a secondary constraint 
\be
\mathcal C_4= \{\pi_{e_0{}^0},\,\int d^3 x\, \mathcal H^T\}\approx 0,
\ee
which does not contain $e_0{}^0$. 
From the expression of the secondary constraint, 
we know it is an equation for the dynamical variables $\g_{ij},\,e_{ij},\,\pi^{ij}$. 
\item
The second key point is that in $\mathcal C_4$ the spatial derivative term is the Riemann curvature tensor of the induced metric $\g_{ij}$. 
This indicates the Poisson bracket of $\mathcal C_4[\a]$ 
and $\mathcal C_4[\b]$ vanishes
\be
\{\mathcal C_4[\a],\,\mathcal C_4[\b]\}=0,
\ee
because $\mathcal C_4$ does not involve $\Pi^{ij}$, 
which is the conjugate momentum of the induced metric. 
\end{itemize}

If $\mathcal C_4$ is a first class constraint, 
it already eliminates 1 dynamical degree of freedom because of the first key point. 

When $\mathcal C_4$ is related to a second class constraint, 
an independent tertiary constraint $\mathcal C_5$ is generated 
by the time derivative of $\mathcal C_4$. 
This tertiary constraint will not contain $e_0{}^0$ due to the second key point. 
The lapse function $N$ and the shift vector $N^i$ are gauge parameters 
associated with the diffeomorphism invariance, 
so they will not appear in the constraint equations
\footnote{However, from the explicit expression of $\mathcal C_5$, 
it is not apparent that $N$ is unconstrained.  
}. 
Then $\mathcal C_5$ is an equation for the dynamical variables
\footnote{In Type B models, the Hamiltonian is quadratic in $e_0{}^i$,  
so the corresponding secondary constraint can be solved 
and $e_0{}^i$ are functions of the dynamical variables.}. 
$\mathcal C_5$ could correspond to a first or second class constraint, 
but in both cases at least 1 degree of freedom is removed 
by the two constraints $\mathcal C_4$ and $\mathcal C_5$. 

The general Hamiltonian structure of these bi-gravity models 
is analogous to Hassan-Rosen bi-gravity theory \cite{Hassan:2011zd} . 
The reason is that both of them are nonlinear completions of the same linear theory, 
Fierz-Pauli massive gravity. 
As a result, the degree of freedom eliminated by $\mathcal C_4$ and $\mathcal C_5$ is a nonlinear Ostrogradsky's scalar ghost, namely the Boulware-Deser ghost. 
\\

We expect that the general constraint structure of the novel two-derivative terms 
can be extended to any dimension 
\be
R\big(E\big)\wedge E\wedge \dots\wedge E
\wedge F\wedge \dots\wedge F
\ee
and to the cases of novel higher-derivative terms
\be
R\big(E\big)\wedge\dots \wedge R\big(E\big)\wedge E\wedge \dots\wedge E
\wedge F\wedge \dots\wedge F. 
\ee 

For multi-gravity generalizations, 
we can introduce different $F^{(k)}$. 
The Boulware-Deser ghost should be absent as well. 
There are additional primary constraints for $\pi^{(k)}_{ij}$ 
because they are functions of $\pounds_n \g_{ij}$. 
To understand the origin of these primary constraints, 
we can eliminate
\be
e_{\m\n}^{(k)}=E_\m{}^AF^{(k)}{}_{\n}{}^B\eta_{AB},
\ee 
by their equations of motion, which is possible in most of the cases. 
Then a multi-gravity theory with novel derivative terms becomes a model of higher curvature gravity, 
so there are at most two dynamical spin-2 fields. 
An interesting question is 
whether the resulting higher derivative gravity model 
is more general than that from a bi-gravity theory.

\begin{acknowledgments}
I would like to thank C. de Rham, A. Matas and A. Tolley for critical comments. 
I am grateful to X. Gao for his participation in the early stages of this project. 
I also thank E. Joung and E. Kiritsis for inspiring comments. 
The tensor computations have been performed with the help of the {\it xAct} package \cite{xAct} for {\it  Mathematica}. 

This work was supported in part by European Union's Seventh Framework Programme
under grant agreements (FP7-REGPOT-2012-2013-1) no 316165, the EU program ``Thales" MIS 375734
 and was also cofinanced by the European Union (European Social Fund, ESF) and Greek national funds through
the Operational Program ``Education and Lifelong Learning" of the National Strategic
Reference Framework (NSRF) under ``Funding of proposals that have received
a positive evaluation in the 3rd and 4th Call of ERC Grant Schemes".
\end{acknowledgments}

\end{document}